\documentclass[aps,prl,nofootinbib,twocolumn,floatfix,notitlepage]{revtex4-1}
\pdfoutput=1

\usepackage{graphicx}
\usepackage{amsmath,amsfonts,amssymb}
\usepackage{color}
\usepackage[breaklinks,colorlinks,urlcolor=blue,citecolor=blue,linkcolor=magenta]{hyperref}
\usepackage{verbatim}
\usepackage{enumitem}
\usepackage{hyperref}
\usepackage{orcidlink}
\usepackage{booktabs}
\usepackage{soul}

\newcommand{\be}{\begin{equation}}
\newcommand{\ee}{\end{equation}}
\newcommand{\bea}{\begin{equation} \begin{aligned}}
\newcommand{\eea}{\end{aligned} \end{equation}}

\newcommand{\Mpl}{M_{\rm pl}}

\newcommand{\td}{{\rm d}}

\begin{document}

\title{Interpreting DESI 2024 BAO: late-time dynamical dark energy or a local effect?}

\author{Ioannis D. Gialamas\orcidlink{0000-0002-2957-5276}}
\email{ioannis.gialamas@kbfi.ee}
\author{Gert H\"utsi\orcidlink{0000-0002-9322-004X}}
\email{gert.hutsi@kbfi.ee}
\author{Kristjan Kannike\orcidlink{0000-0003-3710-317X}}
 \email{kristjan.kannike@cern.ch}
 \author{Antonio Racioppi\orcidlink{0000-0003-4825-0941}}
 \email{antonio.racioppi@kbfi.ee}
 \author{Martti Raidal\orcidlink{0000-0001-7040-9491}}
 \email{martti.raidal@cern.ch}
\author{Martin Vasar\orcidlink{0009-0003-3514-1575}}\email{martin.vasar@ut.ee}
 \author{Hardi Veerm\"ae\orcidlink{0000-0003-1845-1355}}
 \email{hardi.veermae@cern.ch}
 
\affiliation{Keemilise ja Bioloogilise F\"u\"usika Instituut, R\"avala pst. 10, 10143 Tallinn, Estonia}

\begin{abstract}
We perform fits to DESI, CMB and supernova data to understand the physical origin of the DESI hint for dynamical dark energy. We find that the parametrization $w = w_0 + (1 - a) w_a$ can, misleadingly, hint for a phantom Universe, although quintessence models fit the data well. Model-independently, we find that deviations from $\Lambda$CDM are driven by low-$z$ supernova data and take place only at very low redshifts $z < 0.1$. Therefore, either our Universe underwent dramatic changes very recently or we do not fully understand our local Universe in a radius of about $300\,h^{-1}\rm Mpc$. 

\end{abstract}
\maketitle
 
\section{Introduction}

We have come to a turning point where the standard $\Lambda$CDM model, previously regarded as robust, may have to be reconsidered. Taking into account recent measurements of baryon acoustic oscillations (BAO) from the Dark Energy Spectroscopic Instrument (DESI)~\cite{DESI:2024uvr,DESI:2024lzq}, combined with cosmic microwave background (CMB) data~\cite{Planck:2018vyg} and supernova (SN) observations~\cite{DES:2024jxu}, it appears~\cite{DESI:2024mwx} that the $\Lambda$CDM model is losing ground, signalling the onset of a dynamical dark energy (DE) era.

The new DESI data~\cite{DESI:2024mwx} have attracted significant interest from the scientific community, prompting the development of various models that align with these observations. These models include quintessence from scalar fields~\cite{Tada:2024znt,Berghaus:2024kra,Shlivko:2024llw,Bhattacharya:2024hep,Ramadan:2024kmn}, interacting DE~\cite{Giare:2024smz}, dark radiation~\cite{Allali:2024cji}, and DE produced by baryon conversion in cosmologically coupled black holes~\cite{Croker:2024jfg} or in an axion phase transition \cite{Muursepp:2024mbb}. Modifications of gravity in light of the new DESI data have also been discussed~\cite{Yang:2024kdo,Escamilla-Rivera:2024sae}. Additionally, numerous scenarios extending beyond the $\Lambda$CDM model have been proposed~\cite{Yin:2024hba,Gu:2024jhl,Wang:2024qan,Wang:2024hks,Colgain:2024xqj,Wang:2024rjd,Carloni:2024zpl}. A thorough analysis of physically motivated DE models (thawing~\cite{Caldwell:2005tm}, emergent~\cite{Li:2020ybr}, mirage class~\cite{Linder:2007ka}) was presented in~\cite{Lodha:2024upq}. Moreover, the Hubble tension~\cite{Riess:2016jrr, Riess:2021jrx} in light of the new DESI data has been analyzed in~\cite{Clifton:2024mdy,Bousis:2024rnb,Qu:2024lpx,Wang:2024dka,Seto:2024cgo,Pogosian:2024ykm,Jia:2024wix,DES:2024ywx}. The validity of distance duality relation using DESI BAO measurements has been investigated in Ref.~\cite{Favale:2024sdq}, separation of the radial and angular BAO measurements with an attempt to isolate the potentially problematic DESI redshift bins was carried out in Ref.~\cite{Wang:2024pui}. A model-independent analysis has been provided by the DESI team~\cite{DESI:2024aqx}, see also Refs.~\cite{Luongo:2024fww,Mukherjee:2024ryz,Dinda:2024kjf, DES:2024fdw}. The need for evolving DE could also be reached without the new DESI measurements by combining CMB and SN data with older BAO measurements~\cite{Park:2024jns}. Similar hints for evolving DE have also been obtained before the DESI BAO measurements~\cite{Rubin:2023ovl,DES:2024jxu,Colgain:2024ksa}. Recently, also a mild $2-3\sigma$ tension was reported between the sum of neutrino masses expected from oscillation experiments and those derived from cosmological data (CMB, SN, DESI BAO)~\cite{Craig:2024tky,Wang:2024hen}.

The goal of this work is to learn as model-independently as possible which dynamical DE scenarios are preferred by the existing data and which data sets drive the fits. For that purpose, we study two existing parametrizations for the equation of state $w$, the linear parametrization~\cite{Chevallier:2000qy, Linder:2002et}, also used in~\cite{DESI:2024mwx}, and the quintessence parametrization~\cite{Dutta:2008qn, Chiba:2009sj}, and propose a new robust step function parametrization of the equation-of-state parameter $w$, which captures the essential features of the quintessence parametrization while being considerably simpler. 

Our first result is that, when performing global fits to DESI, CMB and SN data, it is important to keep in mind which physical scenario is under study. Using the linear ansatz for $w$, which is motivated by its simplicity but does not represent any physical scenario, gives a misleading impression that the new data prefers phantom DE. Phantom models can be problematic because they violate fundamental principles, such as the null energy condition, and lead to unstable vacuum states due to unbounded Hamiltonians, rendering them incompatible with well-defined particle physics theories. The quintessence and step function parametrizations give comparable or better fits to data without any hint of phantom DE. Thus the preference for a $w<-1$ regime is just an artefact of the choice of parametrization.

The most important result of this work is that the present data indicate that deviations from the constant DE took place very recently. This suggests either a very recent thawing of the DE component in our universe or, alternatively, some effect confined to our local Universe within a radius of about half a Gpc. Moreover, the effect pointing the dynamical DE to low redshifts is dominated by local supernovae -- the tension with the standard $\Lambda$CDM disappears if the low redshift supernovae are not included in the data.

\section{Modelling dynamical dark energy}

We will model DE  with a time-dependent equation of state parameter $w(a)$, 
but with a conserved energy-momentum tensor, that is,
\be
    \partial_t\rho_{\rm DE} + 3 H (1+w) \rho_{\rm DE} = 0\,.
\ee
The energy density of the DE therefore evolves as
\bea\label{eq:rhoDE_evol}
    \rho_{\rm DE}(a)
    &= \rho_{\rm DE, 0} e^{- 3 \int^{a}_{1} \frac{\td a'}{a'} [1+w(a')] }\,, \\
\eea
where $\rho_{\rm DE, 0}$ stands for the present energy density. We will consider the following ans\"atze for $w(a)$.

\paragraph{Linear.} A particularly simple phenomenological parametrization\footnote{In light of DESI data~\cite{DESI:2024mwx}, a polynomial parametrization has been used in~\cite{Yang:2024kdo}, while in~\cite{Escamilla:2024fzq}, the use of trigonometric functions gives rise to an oscillating equation of state.} is given by~\cite{Chevallier:2000qy,Linder:2002et}
\be
    w(a) = w_0 + (1 - a) w_a\,,
\ee
where $w_0$ and $w_a$ are free parameters that determine the variation of $w$ from $w_0$ at the present to $w_0 + w_a$ in the past when $a \to 0$. Due to its simplicity, this parametrization is widely used and was implemented in the recent analysis by DESI collaboration~\cite{DESI:2024mwx} (for an extended discussion of this parametrization and its shortcomings see \emph{e.g.}~\cite{Wolf:2023uno, Cortes:2024lgw}). The dynamics of the DE component is given by
\be
    \rho_{\rm DE}(a)
    = \rho_{\rm DE, 0} a^{- 3 (1+w_0+w_a) } e^{ 3 (a-1) w_a}\, .
\ee
The drawback of this parametrization is that it is not motivated by any particular physical model and is meant to capture the first derivative of $w(a)$ at present. It can violate the null energy condition at high redshifts with $w_0+w_a <-1$ when the data prefers a sharp feature around $z=0$. Thus it is unclear whether this behaviour is a numerical artefact introduced by this particular parametrization or suggests exotic new physics.

\begin{figure*}[t!]
\centering
\includegraphics[width=0.95\textwidth]{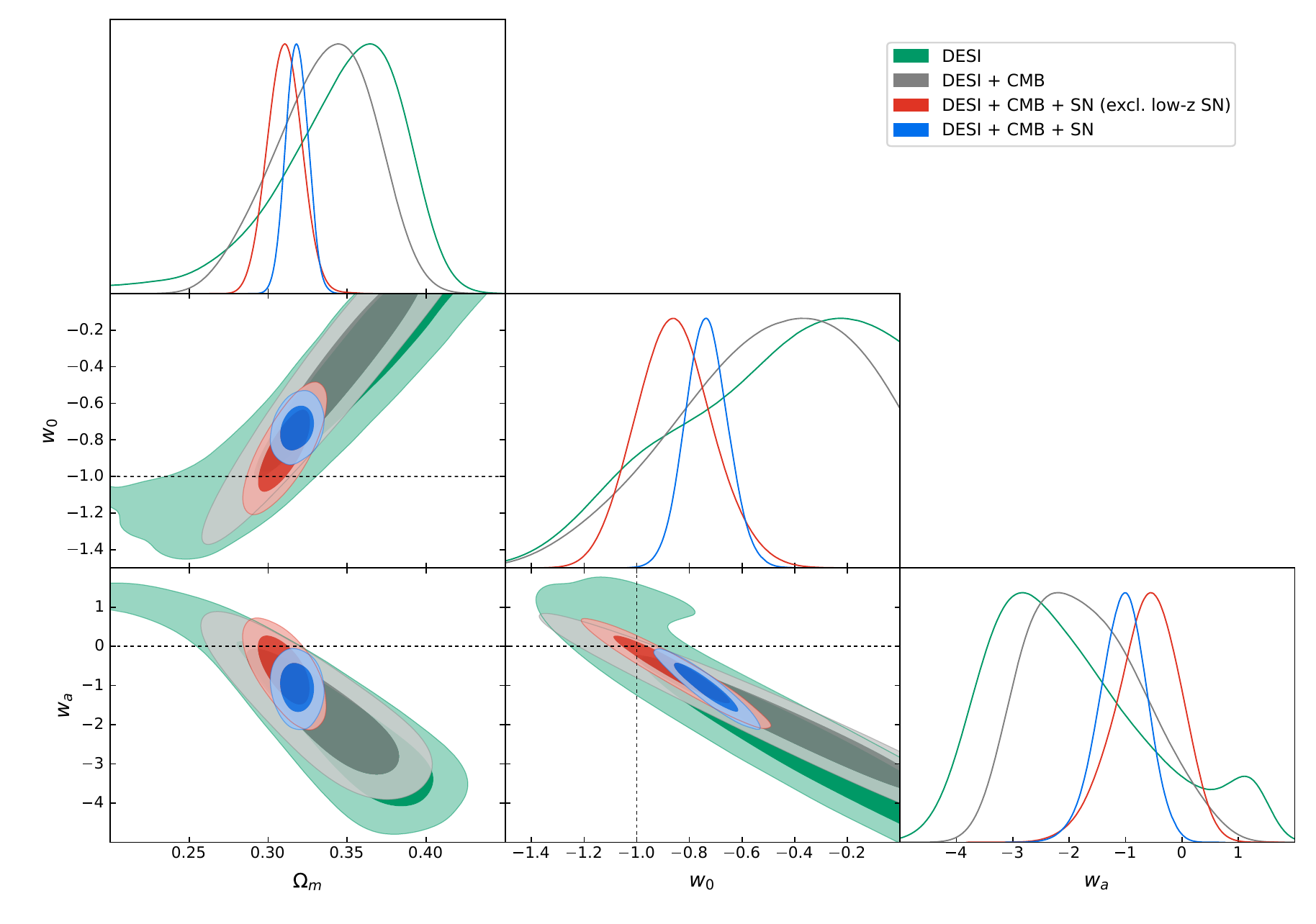}
\caption{Parameter constraints ($1\sigma$ and $2\sigma$ contours) for the \emph{linear} DE model assuming DESI data only, DESI+CMB, and DESI+CMB+SN with and without local SN, as shown in the legend. The locus of the $\Lambda$CDM model is shown by the crossing of dashed lines in the $w_0-w_a$ panel.} 
\label{fig:1}
\end{figure*}

\begin{figure}[t!]
\centering
\includegraphics[width=0.4\textwidth]{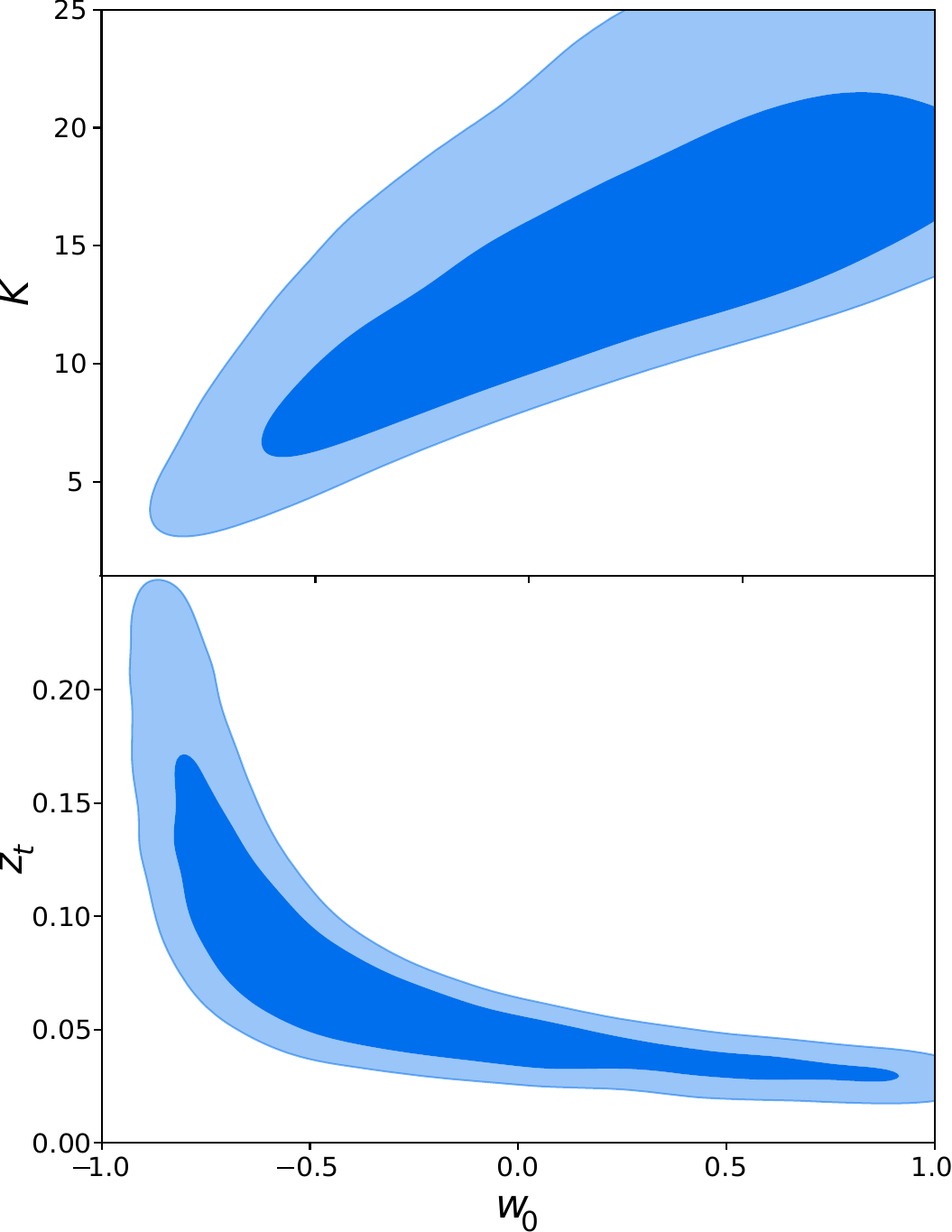}
\caption{The $1 \sigma$ and $2 \sigma$ contours for the quintessence model parameter $K$ (upper panel) and the transition redshift $z_t$ of the step function parametrization (lower panel) against the current $w_0$ parameter, assuming DESI+CMB+SN data.} 
\label{fig:2}
\end{figure}

\begin{figure*}[t!]
\centering
\includegraphics[width=0.32\textwidth]{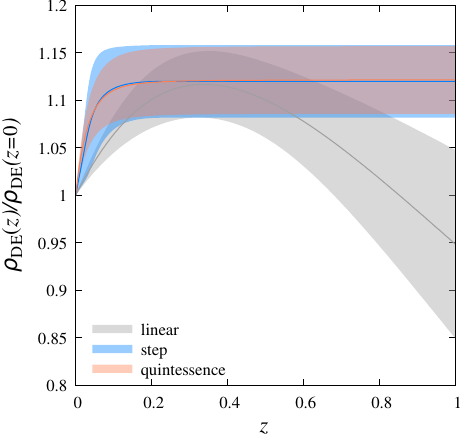}
\includegraphics[width=0.32\textwidth]{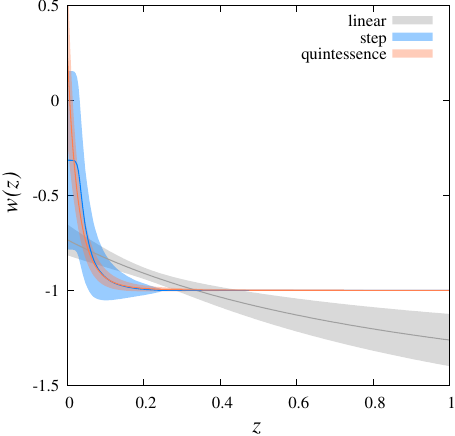}
\includegraphics[width=0.32\textwidth]{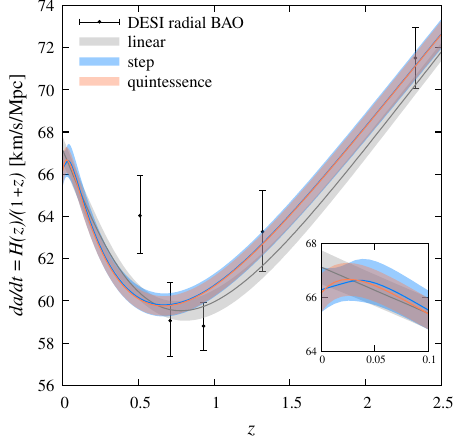}
\caption{Evolution of the posterior mean (with $1\sigma$-scatter) DE density, DE equation of state, and time derivative of the scale factor. The last quantity includes an additional factor of $r_d/r_d^{\rm fid}$ implicitly. Note that the DE equation of state goes slightly phantom for the \emph{step} model. This is caused by the asymmetry of the probability distribution as here we have plotted posterior means along with symmetric $1\sigma$-scatter.} 
\label{fig:3}
\end{figure*}

\paragraph{Quintessence.} The evolution of $w$ in a range of quintessence models (for a review see~\cite{Tsujikawa:2013fta}) with scalar potential $V$ can be approximated by~\cite{Dutta:2008qn,Chiba:2009sj}
\be\label{eq:w:quint}
    w(a)\simeq-1+(1+w_0)a^{3(K-1)}\mathcal{F}(a),
\ee
with
\begin{widetext}
\be
    \mathcal{F}(a)=\left[ \frac{(K-F(a))(F(a)+1)^K+(K+F(a))(F(a)-1)^K}{(K-\Omega_\phi^{-1/2})(\Omega_\phi^{-1/2}+1)^K+(K+\Omega_\phi^{-1/2})(\Omega_\phi^{-1/2}-1)^K}
    \right]^2, \qquad  F(a)=\sqrt{1+(\Omega_\phi^{-1}-1)a^{-3}}\,,
\ee
\end{widetext}
where $\Omega_\phi$ is the current density parameter of $\phi$ and the universe is flat, i.e. $\Omega_\phi+\Omega_\mathrm{m} = 1$. This scenario was also analysed in Ref.~\cite{Tada:2024znt}. The parameter $K$ is determined by the value of the scalar potential $V$ and its second derivative at a given initial field value $\phi_i$, according to the equation $K=(1-4\Mpl^2V''(\phi_i)/3V(\phi_i))^{1/2}$. As a result, depending on the convexity of $V$, $K$ can take any non-negative value, including imaginary ones, driving $w(a)$ to an oscillatory behavior~\cite{Chiba:2009sj,Dutta:2009yb}. However, as shown below, the combined DESI BAO + CMB + SN data favour values of $K\gtrsim 5$.

Although the parametrization~\eqref{eq:w:quint} was initially derived from quintessence models \cite{Dutta:2008qn,Chiba:2009sj}, it can also effectively describe other cosmological models, like $k$-essence models, particularly in scenarios where the field velocity is negligible~\cite{Chiba:2009nh}, and phantom ones, if $w_0 < -1$~\cite{Dutta:2009dr}. Moreover, as we will see, the data indicate a preference for $w_0 > -1$, thereby disfavoring phantom models.

\paragraph{Step.} As a third ansatz we will consider a step function
\be
    w(z) = 
  \begin{cases}
   w_0, & z \leq z_t,
   \\
   w_\infty, & z > z_t, \\
  \end{cases}
\ee
where $w_0$ and $w_\infty$ denote the current and the past equation of state and the transition takes place at redshift $z_t$. We will consider $w_\infty = -1$, in this work. In this model, the evolution of the energy density~\eqref{eq:rhoDE_evol} can be described by a simple analytic relation.

It qualitatively captures the behaviour of thawing quintessence as it models a transition from a time-independent DE energy density into a time-dependent one at $z \leq  z_t$.

\section{Confronting DESI data}

For observational data we use the first year DESI BAO measurements~\cite{DESI:2024mwx}, Planck CMB bounds~\cite{Planck:2018vyg} on angular scale of the sound horizon and supernova luminosity distance measurements from the Dark Energy Survey (DES) Year 5 data release~\cite{DES:2024jxu}. DESI provides BAO measurements at seven redshifts: $z=0.30,0.51,0.71,0.93,1.32,1.49,2.33$. For most of the redshifts anisotropic, \emph{i.e.}, along and perpendicular to line-of-sight measurements of BAO, together with their correlation coefficients are provided, while for redshifts $0.30$ and $1.49$ only isotropized BAO measurements are given. Thus, in total $2\times 5+2=12$ BAO measurements, which are assumed to follow Gaussian error distributions, are used in this work.~\footnote{In our analysis we also implicitly assume that the standard BAO analysis as carried out by the DESI collaboration (which as an input needs assumptions about the fiducial cosmology) is sufficient for our purposes. In case real cosmology deviates significantly from the assumed fiducial model one should rather rely on `purely-geometric BAO' methods as suggested in~\cite{ODwyer:2019rvi,Anselmi:2022exn}. Additionally, it is important to note that this approximate analysis tends to underestimate the statistical uncertainties.} As our analysis does not attempt to achieve absolute calibration of the BAO we work with a combined scaling parameter $r_dh$ ($r_d$ -- sound horizon at the baryon `drag' epoch, $h$ -- reduced Hubble parameter). Thus, \emph{e.g.} the results obtained for the Hubble parameter should be understood as implicitly including an additional factor $r_d/r_d^{\rm fid}$. We consider only spatially flat models\footnote{Including spatial curvature is beyond the scope of this study. However, a similar analysis is possible for non-flat geometries ($\Omega_K \neq 0$)~\cite{Ichikawa:2005nb,Clarkson:2007bc,Anselmi:2022uvj}.} 
and for our fiducial $\Lambda$CDM model (best-fit Planck 2018 $\Lambda$CDM~\cite{Planck:2018vyg}) with $\Omega_mh^2=0.1428$ and $\Omega_bh^2=0.02233$ sound horizon $r_d^{\rm fid}=147.17$~Mpc. Our Markov chain Monte Carlo (MCMC) explorations can be simplified since, \emph{e.g.} marginalizations --  equivalent to maximization under the Gaussianity assumption -- over $r_dh$, which acts as a common multiplicative scaling parameter, can be done analytically.

Instead of the full likelihood calculations, CMB constraints, as in~\cite{Huang:2024qno}, are replaced by a single bound on the angular scale of the sound horizon at recombination inferred from the Planck data: $100\theta_{\rm MC}=1.04090\pm 0.00031$~\cite{Planck:2018vyg}.\footnote{In fact, it turns out that very similar results can also be obtained by assuming a single effective bound on $\Omega_mh^2$, instead.} Although simplified, this captures the main constraint arising from the CMB data.

Supernova measurements are taken from the DES Year 5 data release, which contains 1829 supernovae in total: 1635 genuine DES measurements augmented with 194 local SNIa measurements ($z<0.1$) from the CfA/CSP foundation sample~\cite{Hicken:2009df,Hicken:2012zr,Foley:2017zdq}. Distance moduli, their errors and covariance matrices were taken from the DES GitHub page.\footnote{\url{https://github.com/des-science/DES-SN5YR}} As we are not attempting to achieve absolute magnitude calibration, we are left with an additive normalization constant that needs to be marginalized over, which for the case of assumed Gaussian likelihoods is equivalent to fixing it to its best-fit value. This step can again be done analytically.

Our main results are summarized by Figs.~\ref{fig:1}-\ref{fig:3} and in Table~\ref{table:1}.
Fig.~\ref{fig:1}  is devoted to the \emph{linear} DE model, which has been extensively studied in the literature and as such should serve as a test case for our calculations (corner plots for the \emph{quintessence} and \emph{step function} models can be found in the Appendix). The parameter constraint contours ($1\sigma$ and $2\sigma$) are shown for DESI data alone, DESI+CMB, and DESI+CMB+SN data, the last with and without local supernovae, \emph{i.e.}, including or excluding CfA/CSP sample, respectively.

From Fig.~\ref{fig:1} it is evident that to achieve a reasonable measurement for the DE parameters one needs the input from all the datasets, \emph{e.g.} CMB+DESI data without SN provides only a bound on the linear combination of $w_0$ and $w_a$ (with a best fit, even though this is not particularly meaningful, away from the $\Lambda$CDM point). Similar strongly degenerate measurements, albeit already showing some hints for mild deviations from $\Lambda$CDM, were obtained earlier using SN+CMB data, see \emph{e.g.}~\cite{DES:2024jxu}.

Using the full SN data it is clear that $\Lambda$CDM is disfavored over the linear $w_0-w_a$-parametrization: as shown in Table~\ref{table:1} the respective $\Delta \chi^2$ gets increased by $12.9$, or $\Delta$AIC~\cite{1974ITAC...19..716A} by 8.9.  However, if we remove low-$z$ SN, then the analysis does not require evolving DE and thus fully restores $\Lambda$CDM concordance, the fact most clearly demonstrated by the $\Delta \chi^2$ and $\Delta$AIC values shown in Table~\ref{table:2}. Along similar lines, without the low-$z$ SN, we find that $z_t$ (and $K$) maintain an upper (lower) bound, but the lower (upper) bound, that was present with the low-$z$ SN, vanishes and $\Lambda$CDM (that is, $z_t = 0$ and $K \to \infty$) is now accommodated by the data. Thus it follows that the preference for DE over $\Lambda$CDM is driven by the low-$z$ SN data. 

The constraints for the other two DE models (\emph{step} and \emph{quintessence}) are shown in Fig.~\ref{fig:2}. It turns out that compared to the \emph{linear} model these models allow a much broader range of $w_0$ values, even up to $w_0=1$, corresponding to a fully kinetic energy-dominated scalar field. While the expansion in the best-fit \emph{linear} model still accelerates today, such large values for $w_0$ force the universe to decelerate, as seen in the last panel of Fig.~\ref{fig:3}.

As seen from Table~\ref{table:1}, the goodness of fit for all three DE models is almost equal, with the \emph{step} model performing slightly better than the other two. The preference of evolving DE models over $\Lambda$CDM gets somewhat reduced once penalizing their extra degree of complexity, which according to the Akaike information criterion~\cite{1974ITAC...19..716A} leads to the $\Delta$AIC values as shown in the table.

\begin{table}[t!]
  \centering
  \resizebox{\columnwidth}{!}{%
\small  
  \begin{tabular}{ccccccc}
    \hline 
    \hline \\[-0.4cm]   
    Model & $\chi^2$ & $\Delta \chi^2$ & $\Delta$AIC & $\Omega_m$ & $w_0$ & $w_a\mid z_t\mid K$\\
    \hline\\[-0.4cm] 
    Linear & $1649.7$ & $0.3$ & $0.3$ & [0.31, 0.33] & [-0.89, -0.57] & [-1.92,-0.26]\\[-1pt]
    & & & & {\scriptsize $0.318$} & {\scriptsize $-0.75$} & {\scriptsize $-0.96$} \\
    Step & $1649.4$ & $0.0$ & $0.0$ & [0.31, 0.34] & [-0.93, 0.72] & [0.02,0.19]\\[-1pt]
    & & & & {\scriptsize $0.326$} & {\scriptsize $-0.37$} & {\scriptsize $0.061$} \\
    Quintessence & $1649.8$ & $0.4$ & $0.4$ & [0.31, 0.34] & $w_0 > -0.570$  & [5.2,25]\\[-1pt]
    & & & & {\scriptsize $0.330$} & {\scriptsize $0.53$} & {\scriptsize $15.3$} \\
    \hline
    $\Lambda$CDM & $1662.6$ & $13.2$ & $9.2$ & [0.30, 0.32] & --- & ---\\[-1pt]
    & & & & {\scriptsize $0.310$} && \\
    \hline
  \end{tabular}
  }
  \caption{Goodness of fit and the 95.45\% confidence intervals for the different parametrizations. Best fit values are given in small print below.}
  \label{table:1}
\end{table}

\begin{table}[t!]
  \centering
  \resizebox{\columnwidth}{!}{%
\small  
  \begin{tabular}{ccccccc}
    \hline 
    \hline \\[-0.4cm]   
    Model & $\chi^2$ & $\Delta \chi^2$ & $\Delta$AIC & $\Omega_m$ & $w_0$ & $w_a\mid z_t\mid K$\\
    \hline\\[-0.4cm] 
    Linear & $1465.5$ & $0.0$ & $3.1$ & [0.29, 0.33] & [-1.14, -0.57] & [-1.64, 0.54]\\[-1pt]
    & & & & {\scriptsize $0.306$} & {\scriptsize $-0.90$} & {\scriptsize $-0.38$} \\
    Step & $1466.3$ & $0.8$ & $3.9$ & [0.30, 0.36] & -- & $z_t < 0.18$\\[-1pt]
    & & & & {\scriptsize $0.315$} & {\scriptsize $-0.66$} & {\scriptsize $0.060$} \\
    Quintessence & $1466.3$ & $0.8$ & $3.9$ & [0.30, 0.34] & -- & $K > 7.7$\\[-1pt]
    & & & & {\scriptsize $0.320$} & {\scriptsize $0.56$} & {\scriptsize $24.5$} \\
    \hline
    $\Lambda$CDM & $1466.4$ & $0.9$ & $0.0$ & [0.29, 0.32] & --- & ---\\[-1pt]
    & & & & {\scriptsize $0.304$} && \\
    \hline
  \end{tabular}
  }
  \caption{Same as Table~\ref{table:1} but for the analysis without the local SN data. The $w_0$ confidence intervals are not shown when the posterior is essentially flat.}
  \label{table:2}
\end{table}

The $\chi^2$ values given in Table~\ref{table:1} might look abnormally low for the number of degrees of freedom we have: $1829$~SNe + $12$~BAO measurements + $1$~CMB constraint $-$ $4$~model parameters ($\Omega_m$, $r_dh$, + 2 DE parameters) = 1838.
These low values are fully driven by the SN data. As discussed in Ref.~\cite{DES:2024jxu}, due to the way contamination of the SN data is treated the effective number of SNe is significantly lower $\sim 1735$, rather than $1829$. 

Fig.~\ref{fig:3} shows the evolution of DE density, its equation of state, and the time derivative of the scale factor (the last of which has omitted an implicit factor of $r_d/r_d^{\rm fid}$). It is evident that the only robust and consistent feature amongst the models is the very recent/local ($z\lesssim 0.1$) sharp drop, ${\cal O}(10\%)$, in the DE density  (for similar findings on this topic, see~\cite{Shlivko:2024llw}), or equivalently a sharp rise in the DE equation of state parameter. The fast drop in the DE density and corresponding crossing of the phantom barrier at higher redshifts is an artefact of the $w_0-w_a$-parametrization (quite unavoidable in this setting, since the data demand large negative $w_a$), which is not required by the data.

Similarly, the narrow range of $w_0$ preferred by the linear parametrization can be considered an artefact. This is due to the negative correlation between $w_0$ and $w_a$ shown in Fig.~\ref{fig:1}, and that a very negative $w_a$ suggests strong phantom behaviour already at relatively low redshifts and would thus clash with the DESI BAO data at higher redshifts.

It is instructive to compare our results with a ``model agnostic'' study \cite{DESI:2024aqx}, which uses a relatively low order Chebyshev polynomial basis for $w(a)$. Together with a sharp rise at low $z$, they have a mild preference for phantom behaviour at higher redshifts. Polynomials, however, do not naturally permit rapid changes and thus the high-$z$ behaviour can be dictated by the low-$z$ one, if the latter contains sudden changes in $w$.

\section{Discussion and conclusions}

The findings of our paper can be summed up by the following three points:

\vspace{2mm}
\noindent\emph{ 1) The evolving nature of DE is supported mainly by the low redshift supernova data.}
\vspace{2mm}

In particular, as illustrated in Fig.~\ref{fig:1} (see also Figs.~\ref{fig:4} and~\ref{fig:5} in the Appendix), the tension between the standard $\Lambda$CDM and the combined data is relieved if the low-$z$ supernova data is omitted. 

Contemporary SN samples have become so large that we have entered a regime where the statistical errors can get overwhelmed by systematics. These issues are not easy to handle since SN are complex phenomena whose detailed physics is not well understood. Thus, there are always possibilities for unaccounted-for systematic effects. Also, for nearby SN, the data analysis is more demanding due to the peculiar motions of their host galaxies, which have a significant effect and thus need to be modeled accurately.\footnote{Although our analysis including low-$z$ SN data suggests that deviation from standard $\Lambda$CDM takes place only at low $z$, we cannot rule out that the tension with $\Lambda$CDM might be relieved by considering other data splits that involve excluding data from higher redshifts because we have not considered such possibilities here.}

\vspace{2mm}
\noindent\emph{ 2) Our analysis suggests that, if taken at face value, there is approximately a $10 \%$ decrease in the DE density in our ``local neighbourhood", that is, at redshift $z\lesssim 0.1$.} 
\vspace{2mm}

As the observable volume of the $z\lesssim 0.1$ Universe is limited by causality, it is not possible to tell by direct observation whether this decrease in the energy density is a global effect stemming from a property of DE in the entire Universe or a truly local effect limited to the surrounding $300\,h^{-1}\rm Mpc$. Nevertheless, it is possible to devise indirect model-dependent probes that allow differentiating between the two possibilities. 

As a generic hypothesis consider, for instance, inhomogeneous DE~\cite[see \emph{e.g.}][]{Erickson:2001bq, Bean:2003fb, Weller:2003hw, Hu:2004yd, Hannestad:2005ak, Perivolaropoulos:2014lua} in which case the observations could be explained by a local DE underdensity. This scenario could be tested via the integrated Sachs-Wolfe effect~\cite{Sachs:1967er} and by looking for the imprints DE fluctuations leave on the distribution of matter in our Universe. Similarly, one can consider how structure formation in the local Universe differs in the presence of a time-independent DE underdensity or a time-dependent but uniform DE density.

\vspace{2mm}
\noindent\emph{3) There is no evidence for phantom or quintom behaviour.}
\vspace{2mm}

Potential evidence for phantom behaviour in the $w_0-w_a$ parametrization is an artefact introduced by extrapolating the linear ansatz to larger redshifts when low redshift data prefers a relatively large negative $w_a$. We find that the \emph{quintessence} and the \emph{step} function parametrizations of the equation-of-state parameter employed in our analysis fit the data comparably to the linear $w_0-w_a$-parametrization. By adopting these parametrizations, we offer a more accurate and physically plausible description of DE, eliminating any indications of phantom DE. 

The parameter $w_0$ remains poorly constrained in the \emph{quintessence} and the \emph{step} function parametrizations, with potential values being $w_0\gtrsim -0.6$ and $w_0\gtrsim -0.9$ respectively. The preference for a narrow range of $w_0$ in the \emph{linear} parametrization can be considered another artefact of that ansatz.

To conclude, our fits to data indicate that the present evidence for dynamical dark energy comes mainly from the local supernova data. Therefore, either the fundamental properties of our Universe, characterised by the equation of state $w$ and the Hubble parameter $H$, underwent dramatic changes very recently or, alternatively, we do not fully understand the systematics of our local Universe.

\begin{acknowledgments}
\vspace{5pt}\noindent\emph{Acknowledgments --}
We would like to thank Ville Vaskonen for the discussions. We acknowledge the use of GetDist software~\cite{Lewis:2019xzd}. This work was supported by the Estonian Research Council grants MOB3JD1202, PSG869, PRG803, PRG1055, PRG1677, RVTT3 and RVTT7 and the Center of Excellence program TK202 ``Fundamental Universe''.
\end{acknowledgments}

\vspace{5pt}\noindent\emph{Note added -- } 
During the refereeing process of this paper, the work~\cite{Efstathiou:2024xcq}  appeared whose findings regarding the impact of local SN are in line with our results.

\appendix
\section{Appendix: Corner plots for DE fits}
\label{appendix}

\begin{figure*}[h!]
\centering
\includegraphics[width=\textwidth]{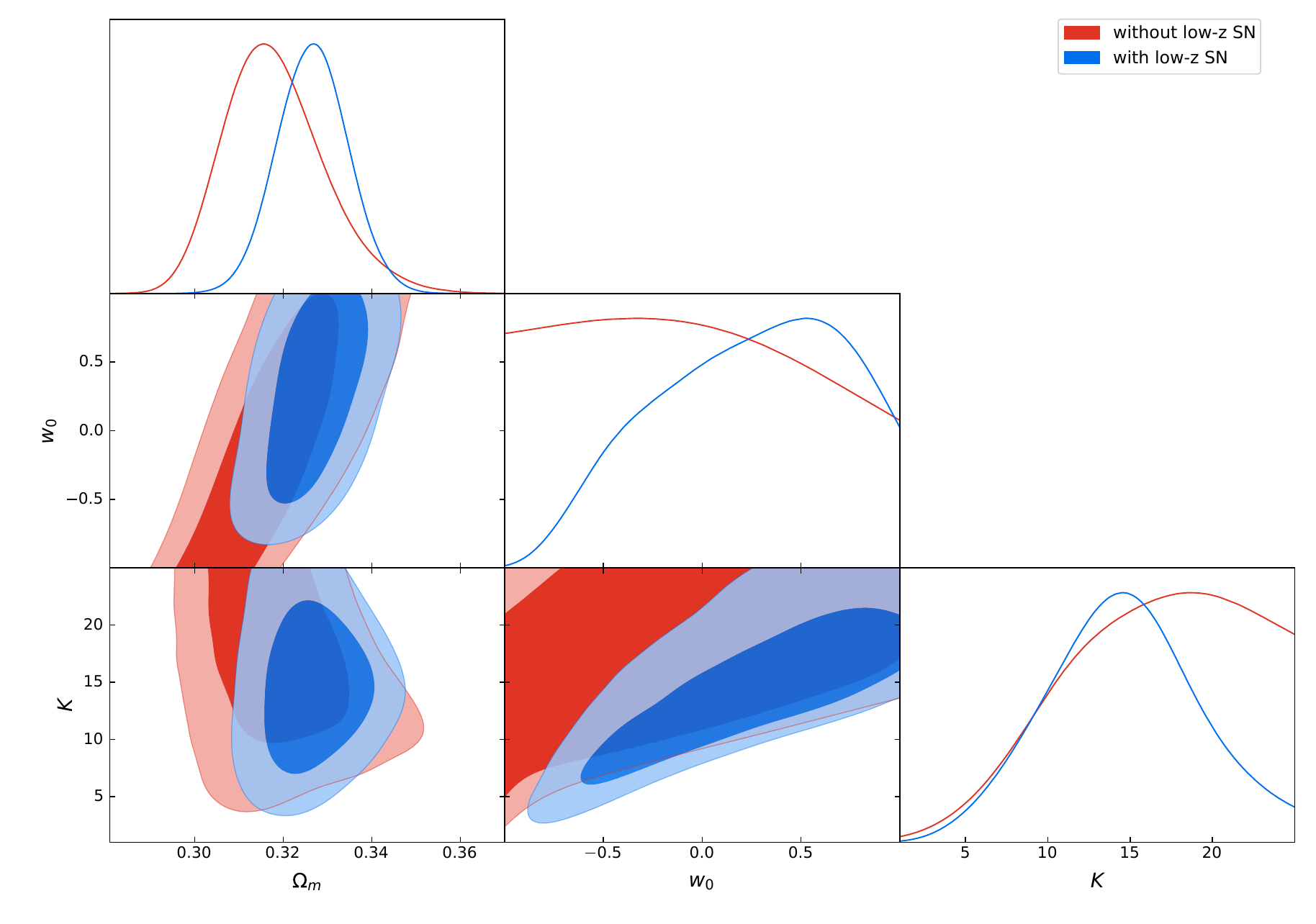}
\caption{Parameter constraints ($1\sigma$ and $2\sigma$ contours) for the \emph{quintessence} DE model assuming DESI+CMB+SN data with and without local SN, as shown in the legend.} 
\label{fig:4}
\end{figure*}

\begin{figure*}[h!]
\centering
\includegraphics[width=\textwidth]{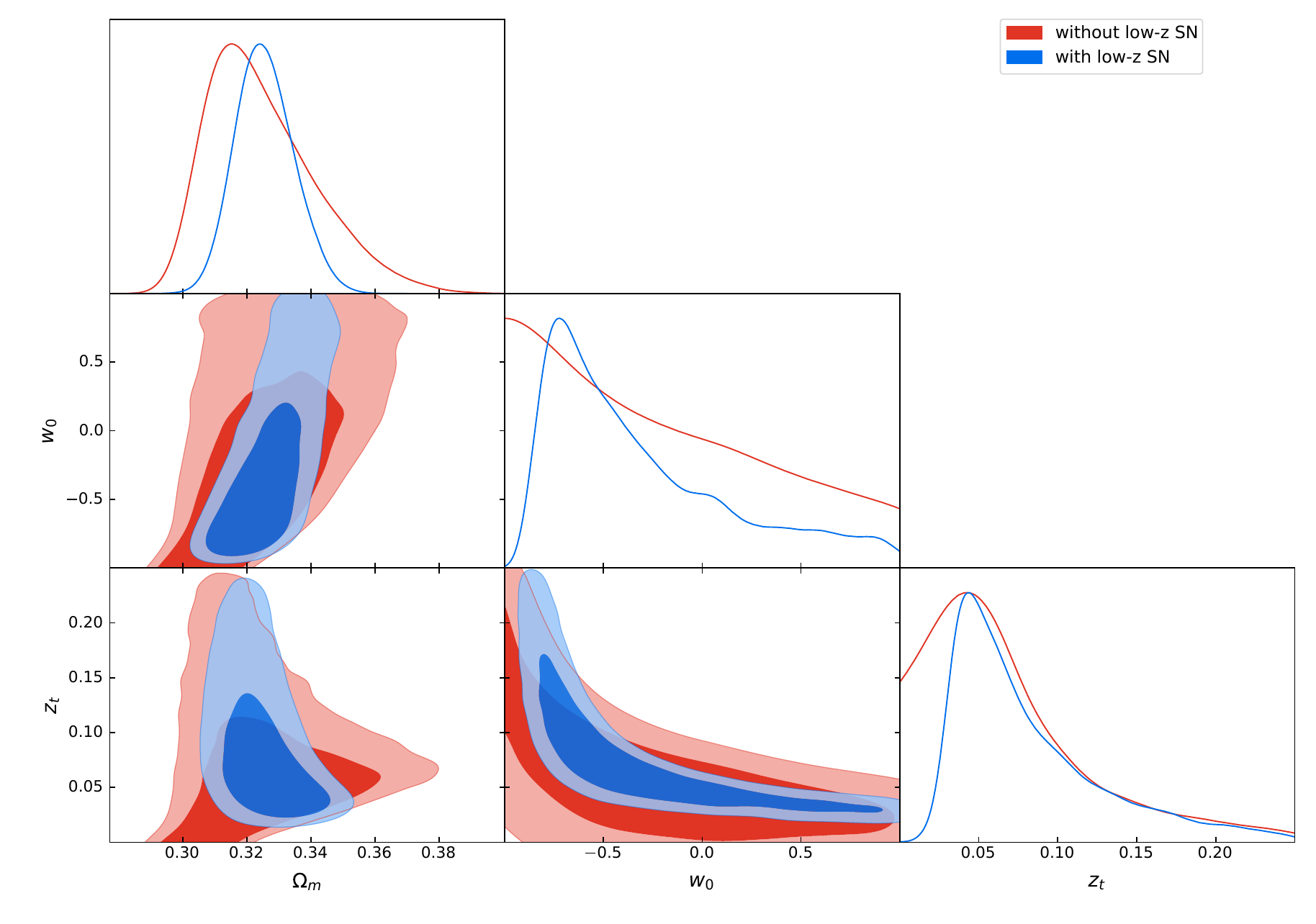}
\caption{Parameter constraints ($1\sigma$ and $2\sigma$ contours) for the \emph{step} DE model assuming DESI+CMB+SN data with and without local SN, as shown in the legend.} 
\label{fig:5}
\end{figure*}

We show here the corresponding corner plots from our likelihood analysis for the \emph{quintessence} (Fig.~\ref{fig:4}) and \emph{step} (Fig.~\ref{fig:5}) models.

\onecolumngrid

\twocolumngrid

\bibliographystyle{JHEP}
\bibliography{dark}

\end{document}